\documentclass[a4paper,12pt]{article}
\pdfoutput=1
\usepackage{epsfig,pdflscape,amssymb}
\usepackage{fancybox,amsmath,color}
\usepackage[vcentermath]{youngtab}

\newcommand{\BR}{\mathbb{R}}

\newcommand{\be}{\begin{equation}}
\newcommand{\bea}{\begin{eqnarray}}

\newcommand{\ee}{\end{equation}}
\newcommand{\eea}{\end{eqnarray}}

\begin{document}

\makeatletter
\@addtoreset{equation}{section}
\makeatother
\renewcommand{\theequation}{\thesection.\arabic{equation}}

\vspace{1.8truecm}

\vspace{15pt}


{\LARGE{ 
\centerline{\bf Central Charges for the Double Coset} 
}}  

\vskip.5cm 

\thispagestyle{empty} 
\centerline{ {\large\bf Shaun de Carvalho${}^{b,}$\footnote{{\tt 542425@students.wits.ac.za}},  
Robert de Mello Koch$^{a,b,}$\footnote{{\tt robert@neo.phys.wits.ac.za}} 
and Minkyoo Kim${}^{b,}$\footnote{{\tt mimkim80@gmail.com}}}}

\vspace{.4cm}
\centerline{{\it ${}^a$ Guangdong Provincial Key Laboratory of Nuclear Science},}
\centerline{{ \it Institute of Quantum Matter, South China Normal University, Guangzhou 510006, China}}

\vspace{.4cm}
\centerline{{\it ${}^b$ National Institute for Theoretical Physics,}}
\centerline{{\it School of Physics and Mandelstam Institute for Theoretical Physics,}}
\centerline{{\it University of the Witwatersrand, Wits, 2050, } }
\centerline{{\it South Africa } }

\vspace{1truecm}

\thispagestyle{empty}

\centerline{\bf ABSTRACT}

\vskip.2cm 

The state space of excited giant graviton brane systems is given by the Gauss graph operators.
After restricting to the $su(2|3)$ sector of the theory, we consider this state space.
Our main result is the decomposition of this state space into irreducible representations of the $su(2|2)\ltimes\BR$ 
global symmetry.
Excitations of the giant graviton branes are charged under a central extension of the global symmetry. 
The central extension generates gauge transformations so that the action of the central extension vanishes on physical states.
Indeed, we explicitly demonstrate that the central charge is set to zero by the Gauss Law of the brane world volume
gauge theory.

\setcounter{page}{0}
\setcounter{tocdepth}{2}

\newpage
\tableofcontents

\setcounter{footnote}{0}

\linespread{1.1}
\parskip 4pt

{}~
{}~

\section{Introduction}\label{intro} 

There is by now exquisite confirmation of the AdS/CFT correspondence \cite{Maldacena:1997re,Gubser:1998bc,Witten:1998qj}.
Many of the precision tests carried out are possible because summing the planar diagrams leads to an integrable model 
for anomalous dimensions of single trace operators which are dual to closed string states \cite{Minahan:2002ve,Beisert:2010jr}. 
The integrable model describes defects (magnons) which are excitations of an infinitely long ``ferromagnetic ground state''.
The ground state preserves half the supersymmetries. 
There are finite size corrections when the chain is finite in length.

The magnon excitations scatter with each other.
A significant insight is that the $S$-matrix of these magnon excitations is completely determined, by symmetry, up to an 
overall phase \cite{Beisert:2005tm,Beisert:2006qh}. 
To simplify the description, consider an infinite spin chain which allows us to study excitations individually. 
The full $PSU(2,2|4)$ symmetry is broken to $SU(2|2)\times SU(2|2)\ltimes \BR$.
Excitations carry the quantum numbers of a central extension of this subalgebra with the central charge measuring 
the quasi-momentum of the excitation\cite{Beisert:2005tm,Beisert:2006qh}. 
The original $PSU(2,2|4)$ does not admit a central extension and for a closed string the net central charge vanishes by level matching constraints.

There are many states in the string theory Hilbert space that are not closed strings.
The theory has D-brane excitations which support open strings.
These D-branes are dual to CFT operators that have a bare dimension of order $N$, so that their large $N$ dynamics 
is not captured by summing planar 
diagrams \cite{Balasubramanian:2001nh,Corley:2001zk,Aharony:2002nd,Berenstein:2004kk,Balasubramanian:2004nb,deMelloKoch:2007rqf,deMelloKoch:2007nbd,Bekker:2007ea}.
In this setting powerful methods based on group representation theory are effective tools with which to attack the large $N$ limit 
\cite{Kimura:2007wy,Brown:2007xh,Bhattacharyya:2008rb,Brown:2008ij,Bhattacharyya:2008xy,Kimura:2008ac}.
A relevant result for us is the diagonalization of the one loop dilatation operator, using a double coset 
ansatz \cite{Koch:2011hb,deMelloKoch:2011ci,deMelloKoch:2012ck}.
This model describes excitations of background branes, with the background branes described using a Young diagram
with long\footnote{Here long means there are order $N$ boxes in the row/column.} rows (for dual giant gravitons)
or columns (for giant gravitons).
The interactions of these excitations have not been explored in much detail 
yet \cite{Koch:2013yaa,Ali:2015yrs,deCarvalho:2018xwx}.
The calculations that are required are technical and quickly become unmanageable.
Given the remarkable success in the planar limit, of a symmetry based approach, it is natural to develop a symmetry
analysis applicable in this setting\footnote{For an early attempt, using a small fraction of the possible symmetries, 
see \cite{Koch:2013xaa}.}. 
The main goal of this paper is to study the $su(2|3)$ sector of the complete theory and show how the global $su(2|2)$ 
symmetry is realized in the resulting Hilbert space of giant graviton branes and their open string excitations.
This result is important since experience from the planar limit suggests that constraints from the global symmetry provide 
powerful insights with which to study excitations of the background branes.
Further, the details are rather intricate so that in the end we arrive at a non-trivial extension of the discussion of
\cite{Beisert:2005tm,Beisert:2006qh}.

The fact that we are considering open strings has some interesting implications, already explored by Berenstein
in \cite{Berenstein:2014zxa}.
Since this discussion is highly relevant for what follows, we will review the key ideas.
To start, consider open superstrings in flat Minkowski spacetime.
The lowest lying string modes of a string stretching between two flat parallel and separated D-branes,  fill out a massive 
short representation of the unbroken supersymmetry of the D-brane system. 
The existence of these representations requires a central charge extension of the unbroken supersymmetry algebra.
The central extension is needed to get a short multiplet.
This additional central charge is an electric charge carried by the string end-points.
Closed string states are not charged so that the central charge is only physical in the open 
string sector or when we compactify the closed string theory on a circle. 
It is measurable in the field theory limit when we spontaneously break the non-abelian gauge symmetry on the stack of branes,
corresponding to the Coulomb branch of the Yang-Mills theory living on the world volume of the D-branes. 

The key conclusion of Berenstein \cite{Berenstein:2014zxa} is that the central charge of the Coulomb branch is a limit 
of the central charge extension of \cite{Beisert:2005tm,Beisert:2006qh}.
Our analysis supports this conclusion.
Note that \cite{Berenstein:2014zxa} is not using the language of the double coset ansatz, but instead employs a
collective coordinate approach \cite{Berenstein:2013md,Berenstein:2013eya,Berenstein:2014isa} which is well suited to 
the semi-classical limit.
Although the collective coordinate and double coset ansatz are rather different descriptions, their conclusions are in
good agreement \cite{Berenstein:2013md}.

In Section \ref{StateSpace} we review the background needed to understand the double coset ansatz. 
Our goal is to provide enough details to develop the Hilbert space of states of the excited giant graviton brane system.
We explain the change of basis from restricted Schur polynomials to Gauss graph operators which are the eigenoperators 
of dilatations.
Gauss graph operators are closely related to the dual gravitational system: they are labeled with a graph that has a vertex for 
each brane in the giant graviton brane system.
The vertices are decorated with directed edges that describe open string excitations.
Our description of the complete state space is novel and in particular we develop the structure of the fermionic 
states which is new.
We then consider the asymptotic symmetries in Section \ref{AsymptoticSymmetries}.
By asymptotic we mean the situation in which impurities are well separated and hence are not interacting.
The discussion is necessarily more complicated than the discussion in \cite{Beisert:2005tm,Beisert:2006qh} because
we have a far bigger space of possible impurities.
The action of the generators of the global symmetry algebra is rather complicated in the restricted Schur polynomial basis.
Reorganizing the basis into irreducible representations of the global symmetry is not trivial.
Remarkably, the basis provided by Gauss graph operators achieves this reorganization! 
Further, excitations again carry a charge under the central extension, echoing what happens in the planar limit.
In the (planar) closed string case the central extension measures the quasi-momentum of the excitations and due to
cyclicity of the trace (which corresponds to level matching in the string description) the total central extension vanishes.
This vanishing of the central extension is necessary, since the algebra on physical states is not centrally extended.
We find an equally compelling description in our non-planar setting.
Giant graviton branes have a compact world volume, so that the Gauss Law constraint of the brane world volume
gauge theory forces the total charge on the world volume to vanish.
This is manifested in the fact that there must must be the same number of directed edges leaving each node as there are
edges terminating on each node.
This condition - which is the requirement that the physical state is gauge invariant - ensures that the total central extension
vanishes.
Further the action of the central charges on the Gauss graph operators has a natural interpretation as a gauge transformation.
We end with some conclusions and discussion in Section \ref{Discussion} including speculations on how the global
symmetry might be used to study interactions between excitations.
The Appendices collect technical details that are used to develop the arguments of the paper.

\section{State space}\label{StateSpace}

The operators we consider are built from three complex bosonic matrices $X,Y,Z$ and two complex fermionic 
matrices $\psi_1,\psi_2$.
These fields all transform in the adjoint of the $U(N)$ gauge group.
This sector of the theory is a closed subsector and it enjoys an $su(2|3)$ supergroup global symmetry.
We will construct the branes in our giant graviton brane system using only the $Z$ field.
The brane system without excitations is a $1/2$ BPS operator.
A linear basis for the brane system without excitations is provided by the Schur polynomials, which are labeled by
a single Young diagram.
Each giant graviton brane corresponds to a long column and each dual giant graviton to a long row. 
Excitations are described using $X,Y$ and $\psi_1,\psi_2$.
Generic excited brane states do not preserve any supersymmetry.
A linear basis for the excited brane system is provided by the restricted Schur polynomials, which has a number of Young
diagram labels (one for each type of field and one for the entire collection) as well as multiplicity labels.
The global $su(2|3)$ symmetry of this subsector is not very useful as it relates operators with different numbers of excitations.
For this reason, following \cite{Beisert:2003ys}, we will restrict our attention to the $su(2|2)$ subgroup which does
preserve the number of excitations.
In this section we will give a complete description of the excited giant graviton brane state space that will be organized,
in the next Section, by the global $su(2|2)$ symmetry. 

\subsection{Restricted Schur Polynomials}

The restricted Schur polynomials provide a linear basis\footnote{Here by linear basis we simply mean that any local gauge 
invariant operator can be expressed as a sum of restricted Schur polynomials. There is never a need, for example, to square
a restricted Schur polynomial.} for the gauge invariant operators of a generic multi-matrix model.
They correctly account for all constraints following from cyclicity or finite $N$ (trace) relations.

In what follows, we use $b^{(0)}$ to denote the number of $Z$ fields.
Consequently, $b^{(0)}=O(N)$.
We also use $b^{(1)},b^{(2)},f^{(1)}$ and $f^{(2)}$ to denote the number of $Y,X,\psi_1$ and $\psi_2$ fields respectively.
The integers $b^{(1)},b^{(2)},f^{(1)},f^{(2)}$ are at most $O(\sqrt{N})$. 
The total number of fields is denoted $n_T=b^{(0)}+b^{(1)}+b^{(2)}+f^{(1)}+f^{(2)}$.

A restricted Schur polynomial is constructed by tracing a projection operator with the multi-linear operator constructed
from a tensor product of matrices.
The projection operator projects both the collection of row indices and the collection of column indices, onto a definite
representation of $U(N)$, and therefore, by Schur-Weyl duality, onto a definite representation of the permutation group 
which permutes indices of different fields.
The projector first places the complete set of $n_T$ indices into a definite representation, labeled by Young diagram $R$
with $n_T$ boxes.
It then places each of the $b^{(i)}$ indices, for each species of bosonic field, into a definite representation labeled
by a Young diagram $b_i$, which has $b^{(i)}$ boxes.
Finally, it places the $f^{(i)}$ row indices of each fermion species into the representation $f_i$ and the column indices into 
the representation $f_i^T$, each of which have $f^{(i)}$ boxes.
$s^T$ is obtained from $s$ by flipping the Young diagram so that rows and columns are exchanged.  
The reason why bosonic row and column indices are placed into the same representation, is so that the trivial representation
of the symmetric group (labeled by a Young diagram with a single row) appears in the tensor product of row and column
indices.
The trace projects to this trivial representation which is necessary since it follows from bosonic statistics.
Further, the reason why fermionic row and column indices are projected as they are, is so that the antisymmetric
representation of the symmetric group (labeled by a Young diagram with a single column) appears in the tensor product 
of row and column indices.
The trace projects to this antisymmetric representation which is necessary since it follows from fermionic statistics.
For a technical derivation of these facts see \cite{Koch:2012sf,Berenstein:2019esh}.
Thus, $R$ is an irreducible representation of $S_{n_T}$, while the collection of five Young diagrams
$(\{b_i\},\{f_i\})$ label an irreducible representation of the subgroup 
$S_{b^{(0)}}\times S_{b^{(1)}}\times S_{b^{(2)}}\times S_{f^{(1)}}\times S_{f^{(2)}}\subset S_{n_T}$.
The representation $(\{b_i\},\{f_i\})$ of the subgroup may appear more than once upon restricting the representation $R$
of the group.
For that reason we need multiplicity labels.
Following the construction presented in \cite{Koch:2012sf}, we need a label for each of the four Young 
diagrams $b_1,b_2,f_1,f_2$.
The Young diagram $b_0$ appears without multiplicity.
We write these multiplicity labels as a vector $\vec\mu$. 
To get a non-zero trace, the Young diagram labels for the row and column indices must match as explained above.
Multiplicity labels can differ.
Consequently we can write the restricted Schur polynomials as $\chi_{R,(\{b_i\},\{f_i\})\vec{\mu}_r\vec{\mu}_c}$.
Rescaling to produce an operator with unit two point function we obtain $O_{R,(\{b_i\},\{f_i\})\vec{\mu}_r\vec{\mu}_c}$.
In what follows, any operator denoted with a capital letter $O$ has been rescaled so that it has a unit two point function.

A useful approach towards the construction of the restricted Schur polynomial entails starting with $R$ and then
peeling off $f^{(i)}$ boxes, which are then reassembled to produce $f_i$ with multiplicity labels, and then peeling
$b^{(i)}$ boxes, which are then reassembled to produce $b_i$. After peeling off 
$f^{(1)}+f^{(2)}+b^{(1)}+b^{(2)}$ boxes from $R$ we are left with $b_0$.
This makes it clear that $b_0$ appears without multiplicity and that the excitations live at the right most corners of 
$R$, something we will need below.  
Further, it is clear that every box in the Young diagram $R$ is associated with a definite species of field.

Any multitrace operator can be written as a linear combination of restricted Schur polynomials.
In the free field theory limit, the two point function boils down to computing the trace of a product of two projection operators.
This can be done exactly and one finds that the restricted Schur polynomials diagonalize the free field two point function.
Finally, the finite $N$ (trace) relations are simply recovered as the statement that the restricted Schur polynomial vanishes
whenever any of the Young diagrams labeling the polynomial has more than $N$ rows.

A key fact that we will need below to understand the state space of the excited brane system, concerns the number of values 
a pair of multiplicity labels $\vec\mu_r,\vec\mu_c$ can take.
This is expressed in terms of the Littlewood-Richardson number $g(r_1,\cdots,r_k;R)$ which is a non-negative integer counting
how many times $U(N)$ representation $R$ appears in the tensor product $r_1\otimes \cdots \otimes r_k$ of $U(N)$
representations.
For the restricted Schur polynomial $\chi_{R,(\{b_i\},\{f_i\})\vec{\mu}_r\vec{\mu}_c}$ 
we find that $\vec\mu_r,\vec\mu_c$ takes 
\bea
g(b_0,b_1,b_2,f_1,f_2;R)g(b_0,b_1,b_2,f_1^T,f_2^T;R)\label{GenCount}
\eea
values \cite{Koch:2012sf}.
Since the Littlewood-Richardson number also counts the multiplicity of representations of the symmetric group after
restriction\cite{FH}, this formula is not too surprising. 

Our discussion in the subsection above aims to give the reader an understanding of the labels of the restricted Schur polynomials.
This is essentially all we use below.
For a detailed technical derivation of the results reviewed the reader should consult 
\cite{Bhattacharyya:2008rb,Bhattacharyya:2008xy,Koch:2012sf}.

\subsection{Double Coset Ansatz}

The restricted Schur polynomials do not have a definite scaling dimension.
However, they only mix weakly under the action of the dilatation operator: at order $g_{YM}^{2L}$
it is possible for two operators to mix if and only if they differ at most by moving $L$ boxes in any of their
Young diagram labels \cite{DeComarmond:2010ie,Koch:2011hb}.
We want to solve the mixing problem which amounts to finding linear combinations of restricted Schur polynomials
that are eigenoperators of the dilatation operator, and finding their eigenvalues.
There is a limit in which the mixing problem simplifies dramatically.
Recall from the previous section that excitations are located at the right hand corners of the Young diagram $R$.
We expect that the excitations are essentially free if they are well separated, which leads to the displaced corners
approximation \cite{Carlson:2011hy,Koch:2011hb}.
The displaced corners approximation holds for a specific shape of the Young diagram $R$.
Imagine that $R$ has order 1 long rows. 
Starting from the right most box in any row of $R$ and moving to the right most box in any other row, along
the shortest path in the Young diagram $R$, if we always need to move through $O(N)$ boxes, then the displaced
corners approximation can be used.
In the displaced corners approximation there is major simplification in the action of the symmetric group: permutations
acting on the impurities simply swap the boxes associated to the excitation.
Without the displaced corners approximation, the result of a permutation is a linear combination of the original state
and the state with the impurities swapped \cite{Carlson:2011hy,Koch:2011hb}.
This simplified action has two important consequences:
\begin{itemize}
\item[1.] There is a new symmetry: restricted Schur polynomials are invariant (up to a sign - for fermions) 
under swapping impurities that belong to a given row. There is an independent symmetry for the row and column indices.
\item[2.] This symmetry results in a new ``conservation law'': restricted Schur polynomials can only mix if they have
the same number and type of excitations in each row. Consequently the number of each species of excitation in each row is
conserved\cite{Koch:2011hb}.
\end{itemize}
This conservation law holds only at the leading order at large $N$.
There is a compelling physical interpretation of the new conservation law: each row in $R$ is identified with a giant
graviton brane.
Identifying the excitations as open strings we have recovered the statement that Chan-Paton factors are conserved
at zero string coupling.

The mixing problem can be solved by making maximal use of the extra symmetry present in the displaced corners approximation.
Let $H$ denote the permutation group that swaps indices of excitations belonging to the same row.
Another copy of the same group will swap indices of excitations belonging to the same column.
$H$ is a product of symmetric groups, one for each excitation species and for each row (or column) of $R$.
The group of permutations acting on the impurities is given by 
$S_{\rm exc}=S_{b^{(1)}}\times S_{b^{(2)}}\times S_{f^{(1)}}\times S_{f^{(1)}}$.
The extra symmetry implies that we have an operator for each element in the double coset
\bea
H \setminus S_{\rm exc}/ H\label{dcoset}
\eea
The elements of this double coset correspond to graphs, with vertices representing branes (one for each row of $R$) and 
directed edges representing oriented strings (one for each excitation field).
We will sometimes draw one graph for each species of excitation to unclutter the description.
The graphs can be described using some numbers. 
Focus on a single species of excitation and imagine there are a total of $m$ excitations of this species and that 
$R$ has $p$ rows.  
Each excitation corresponds to an edge.
Divide each edge into two halves and label each half.
Use the orientation of the edges to distinguish out going and in going ends and label the out going ends with numbers
$\{ 1, \cdots , m \} $ and the in going ends with the same numbers. 
It is natural to specify how the halves are joined by a permutation $\sigma \in S_m$. 
Let $(m_1, m_2, \cdots , m_p)$ record the number of excitations in each row of $R$ so that $m_1 +m_2 + \cdots m_p = m$. 
By the Gauss law, the numbers of edges leaving or ending at each vertex are given by the same ordered sequence of integers 
$(m_1, m_2 , \cdots , m_p)$. 
Choose the labels of the half-edges such that the ones emanating from the first vertex are labeled $\{1,2,\cdots ,m_1\}$, 
those emanating from second vertex are labeled $\{ m_1 +1 , \cdots  m_1+m_2 \} $ and so on.  
Likewise the half-edges incident on the first vertex are labeled $\{1,2,\cdots,m_1\}$, those incident on the second vertex 
are labeled $\{m_1+1,\cdots m_1+m_2\}$ etc. 
The structure of the graph is specified by the permutation $\sigma \in S_m$ which describes how the $m$ out going half-edges
are joined with the $m$ in going half-edges.
A single graph corresponds to many possible permutations because the $m_i$ strings emanating from  the $i$'th vertex 
are indistinguishable, as are the $m_i$ strings terminating on the $i$'th vertex. 
Thus permutations which differ only by swapping end points that connect to the same vertex do not describe 
distinct configurations.
This symmetry group is nothing but the group $H$ introduced above which makes it clear why the double coset
(\ref{dcoset}) describes the space of restricted Schur polynomials in the displaced corners limit.

The most direct and natural use of the double coset which appears above, is through a Fourier transform.
Remarkably, it turns out that the Fourier transform of the restricted Schur polynomial defines an eigenoperator of the 
dilatation operator \cite{deMelloKoch:2012ck}.
The transformation from the restricted Schur polynomials to the Gauss graph operators replaces the Young diagram and
multiplicity labels for each species of excitation with a permutation $\sigma$.
Consequently, since the transformation works separately for each species, we can simplify the discussion and focus on a 
single species at a time.
The transformation for bosonic excitations was worked out in \cite{deMelloKoch:2012ck} and is as follows
\bea
  O_{R,r}(\sigma)=\sum_{s\vdash m}\sum_{\mu_1,\mu_2}C^{(s)}_{\mu_1\mu_2}(\sigma) O_{R,(r,s)\mu_1\mu_2}
  \label{bosonicgaussgraphoperators}
\eea
Here our bosonic excitation is organized by Young diagram $s$ with multiplicity labels $\mu_1,\mu_2$ in the restricted
Schur basis. 
After transformation, the state of the excitations is described by permutation $\sigma$.
Denote the matrix representing $\tau\in S_m$, in the irreducible representation labeled by Young diagram $s$, by 
$\Gamma^s(\tau)$. 
The transformation coefficient is given by
\bea
   C^{(s)}_{\mu_1\mu_2}(\tau)=|H|\sqrt{d_s\over m!}\sum_{k,m=1}^{d_s}\left(\Gamma^{s}(\tau)\right)_{km}
                                 B^{s\to 1_H}_{k\mu_1}B^{s\to 1_H}_{m\mu_2}
\eea
where we have made use of the branching coefficient defined by
\bea
  \sum_{\mu}B^{s\to 1_H}_{k\mu}B^{s\to 1_H}_{l\mu} = {1\over |H|}\sum_{\gamma\in H}\Gamma^{s}(\gamma)_{kl}
\eea
and $d_s$ is the dimension of irreducible representation $s$.
The branching coefficients $B^{s\to 1_H}_{l\mu}$ resolve the multiplicities that arise when we restrict irrep $s$ of $S_m$
to the identity representation $1_H$ of $H$ for which $\Gamma^{1_H}(\gamma)=1$ $\forall\gamma$.  
The transformation for fermionic excitations was worked out in \cite{Koch:2012sf} and is as follows
\bea
  O_{R,r}(\sigma)=\sum_{s\vdash m}\sum_{\mu_1,\mu_2}\tilde{C}^{(s)}_{\mu_1\mu_2}(\sigma) O_{R,(r,s)\mu_1\mu_2}
  \label{fermionicgaussgraphoperators}
\eea
where the transformation coefficient is given by
\bea
   \tilde{C}^{(s)}_{\mu_1\mu_2}(\tau)=|H|\sqrt{d_s\over m!}\sum_{k,m=1}^{d_s}
\left(\Gamma^{s}(\tau)\hat{O}\right)_{km}
                                 B^{s\to 1_H}_{k\mu_1}B^{s^T\to 1^m}_{m\mu_2}
   \label{fermiontransformation}
\eea
where we have made use of the branching coefficient defined by
\bea
  \sum_{\mu}B^{s^T\to 1^m}_{k\mu}B^{s^T\to 1^m}_{l\mu} = {1\over |H|}\sum_{\gamma\in H}{\rm sgn}(\gamma)\Gamma^{s^T}(\gamma)_{kl}
\eea
The branching coefficients $B^{s^T\to 1^m}_{l\mu}$ resolve the multiplicities that arise when we restrict irrep $s^T$ of $S_m$
to the representation $1^m$ of $H$ for which $\Gamma^{1^m} (\gamma)={\rm sgn}(\gamma)$ $\forall\gamma$. 
Here ${\rm sgn}(\gamma)$ is the sign of the permutation $\sigma$. 
The operator $\hat{O}$ appearing in (\ref{fermiontransformation}) is defined by
\bea
  \hat{O}_{jl}=S^{[1^n]\, s\, s^T}_{\,\qquad j\, l}
\eea
where $S^{[1^n]\, s\, s^T}_{\,\qquad j\, l}$ is the Clebsch-Gordon coefficient, moving between states in the tensor
product $s\times s^T$ and the state spanning $1^m$.
To get some feeling for $\hat{O}$ note that it satisfies
\bea
  \Gamma^{s}_{ij}(\sigma)\hat{O}_{jp}= {\rm sgn} (\sigma)\hat{O}_{ik}\Gamma^{s^T}_{kp}(\sigma)
\eea
and hence $\hat{O}_{jl}$ is a map from $s^T$ to $s$. 
$\hat{O}^T\hat{O}$ maps from $s^T$ to $s^T$ and it commutes with all elements of the group. 
Thus, by Schur's Lemma, it is proportional to the identity.
$\hat{O}\hat{O}^T$ maps from $s$ to $s$ and it commutes with all elements of the group. 
Thus it is also proportional to the identity.
By normalizing correctly we can choose
\bea
   \hat{O}^T\hat{O}={\bf 1}_{s^T}\qquad \hat{O}\hat{O}^T={\bf 1}_{s}
\eea
We use transformation formulas (\ref{bosonicgaussgraphoperators}) and (\ref{fermionicgaussgraphoperators}) below.
See Appendix \ref{GaussGraphTransformations} for technical details of how to applying these transformations.

The Gauss graph operators we consider can have all four species of excitations participating.
The operator is written as $O^{b^{(1)},b^{(2)},f^{(1)},f^{(2)}}_{R,b_0}(\sigma)$.
If it is clear from context, we suppress the $b^{(1)},b^{(2)},f^{(1)},f^{(2)}$ superscript.
The permutation $\sigma\in S_{\rm exc}$ describes how half edges for all excitations are joined. 
As mentioned above, these operators have a good scaling dimension.
From formula (2.1) of \cite{tseytlin}, or $H_2$ of Table 1 of \cite{Beisert:2003ys}, we have the 
one loop dilatation operator 
\bea
  D=&-&g_{YM}^2 \left(\sum_{i>j=1}^3 \, {\rm Tr}\left(\left[\phi_i,\phi_j\right]\left[\partial_{\phi_i},\partial_{\phi_j}\right]\right)
     +\sum_{i=1}^3\sum_{a=1}^2 \, {\rm Tr}\left(\left[\phi_i,\psi_a\right]\left[\partial_{\phi_i},\partial_{\psi_a}\right]\right)\right.
\cr
&+&\,{\rm Tr}\left(\left\{\psi_1,\psi_2\right\}\left\{\partial_{\psi_1},\partial_{\psi_2}\right\}\right)
\Bigg)
\label{fullD}
\eea
where $\phi_i$ has $i=1,2,3$ and stands for $Z,Y,X$.
Since the number of excitations is much smaller than the number of $Z$ fields, interactions between excitations is
subleading and we can work with the simplified expression
\bea
  D&=&-g_{YM}^2\Bigg( {\rm Tr}\left(\left[Z,Y\right]\left[\partial_{Z},\partial_{Y}\right]\right)
                             +{\rm Tr}\left(\left[Z,X\right]\left[\partial_{Z},\partial_{X}\right]\right)\cr
   &&\qquad  +\left. \sum_{a=1}^2 \, {\rm Tr}\left(\left[Z,\psi_a\right]\left[\partial_{Z},\partial_{\psi_a}\right]\right)\right)
\label{simpleD}
\eea
The action of the dilatation operator on this Gauss graph operator is given by
\bea
D O_{R,r}(\sigma_1)  =
   - g_{YM}^2 \sum_{i<j}   ~~ n_{ij} (  \sigma_1  )
 \Delta_{ij}  O_{R,r}(\sigma_1)
 \label{lovelyanswer}
\eea
where $\Delta_{ij}$ acts only on Young diagrams $R,r$.
The integer $n_{ij}$ counts the total number of directed edges (both directions counted) stretched between nodes $i$ and $j$.
The operator $\Delta_{ij}$ splits into three terms
\bea
  \Delta_{ij}=\Delta_{ij}^{+}+\Delta_{ij}^{0}+\Delta_{ij}^{-}
\eea
To describe the action of these three pieces, we need some notation.
Denote the row lengths of $r$ by $r_i$. 
Young diagram $r_{ij}^+$ is obtained by removing a box from row $j$ and adding it to row $i$ and $r_{ij}^-$ is 
obtained by removing a box from row $i$ and adding it to row $j$.
See Appendix \ref{YoungDiagramNotations} for examples of this notation.
We now have
\bea
  \Delta_{ij}^{0}O_{R,(r,s)\mu_1\mu_2} = -(2N+r_i+r_j)O_{R,(r,s)\mu_1\mu_2}
  \label{0term}
\eea
\bea
  \Delta_{ij}^{+}O_{R,(r,s)\mu_1\mu_2} = \sqrt{(N+r_i)(N+r_j)}O_{R^+_{ij},(r^+_{ij},s)\mu_1\mu_2}
  \label{pterm}
\eea
\bea
  \Delta_{ij}^{-}O_{R,(r,s)\mu_1\mu_2} = \sqrt{(N+r_i)(N+r_j)}O_{R^-_{ij},(r^-_{ij},s)\mu_1\mu_2}
  \label{mterm}
\eea
Note that $R$ and $r$ change in exactly the same way so that the number of excitations in each row is preserved by 
the dilatation operator. 
The operators of definite scaling dimension now follow by diagonalizing the action of $\Delta_{ij}$. This problem
was studied in detail in \cite{deMelloKoch:2011ci,Lin:2014yaa}, where in a suitable scaling limit, the problem 
was reduced to the diagonalization of decoupled oscillators.

\subsection{Bosonic State Space}

To specify the states of the $Y$ and $X$ excitations, specify the permutation that joins the half edges
of these excitations, or equivalently, give the graph that the permutation describes.
For the sake of clarity, draw the $X$ and $Y$ edges as separate graphs.
The reader should bear in mind that corresponding nodes are identified, since they correspond to the same row in $R$. 
The $X$ and $Y$ impurities populate neighboring boxes in $R$.
There is a distinct (orthogonal) state for each choice of the pair of Young diagrams $R$ and $r$ and the $X$ and $Y$ graphs.
The rules for drawing a valid graph for a given excitation species are 
\begin{itemize}
\item[1.] There is a graph for each type of excitation. 
The nodes in the graph correspond to the rows in $R$.
Each excitation field appearing in the operator corresponds to a directed edge in the graph.
There is no upper limit on the number of edges.
\item[2.] The number of edges emanating from a given node is equal to the number of edges terminating on the node
which is also equal to the number of excitation boxes (of the given species) in the corresponding row of $R$.
\end{itemize}

\subsection{Fermionic State Space}\label{FermionicStateSpace}

There is an additional rule that must be applied when drawing the graphs for fermionic excitations. 
To motivate the rule, consider the simplest case in which we have $\psi_1$ excitations, but no $X,Y$ or $\psi_2$ excitations.
We can simplify the general counting formula appearing in (\ref{GenCount}) to
\bea
n_{\rm graphs}=g(b_0,f_1;R)g(b_0,f_1^T;R)
\eea
If we have a single excitation $f_1=f_1^T={\tiny\yng(1)}$.
In this case, $n_{\rm graphs}=1$ and we simply have a closed loop on the node corresponding to the row from
with a box is removed from $R$ to produce $b_0$.
Now, imagine removing two impurities from a single row.
In this case we have $f_1={\tiny\yng(2)}$ and $f_1^T={\tiny\yng(1,1)}$ and we find
\bea
g(b_0,f_1;R)=1\qquad\qquad g(b_0,f_1^T;R)=0
\eea
so that there is no restricted Schur polynomial and $n_{\rm graphs}=0$. 
If we have two fermionic excitations, they can't be removed from the same row.
Removing the two excitations from two distinct rows and again taking $f_1={\tiny\yng(2)}$ and $f_1^T={\tiny\yng(1,1)}$
we find
\bea
g(b_0,f_1;R)=1\qquad\qquad g(b_0,f_1^T;R)=1
\eea
We could also have taken $f_1={\tiny\yng(1,1)}$ and $f_1^T={\tiny\yng(2)}$, so that there are two Gauss graph operators
that can be defined.
A little work (see Appendix \ref{GaussGraphTransformations} for useful details) shows that the resulting graphs have
two edges, with opposite orientation either (i) stretched between the two nodes or (ii) forming closed loops on each node.
If we remove three excitations, two from a single row and then the third from a distinct row, we find that there are
two possibilities.
First, $s={\tiny \yng(3)}$ and $s^T={\tiny \yng(1,1,1)}$, or second  $s={\tiny \yng(2,1)}=s^T$.
It is simple to demonstrate that
\bea
g(b_0,{\tiny\yng(3)};R)=1\qquad\qquad g(b_0,{\tiny\yng(1,1,1)};R)=0
\eea
so that the first possibility does not lead to a restricted Schur polynomial and hence no Gauss graph operator.
For the third possibility we have
\bea
g(b_0,{\tiny\yng(2,1)};R)=1
\eea
so that we can define a singe Gauss graph operator.
In Appendix \ref{GaussGraphTransformations} we show that the resulting graph has three edges.
there is a closed loop attached to the node corresponding to the row with two impurities removed, 
as well as two edges with opposite orientation, stretched between the two nodes.
Motivated by the above examples, we have found a simple rule that explains which fermion graphs are possible:
\begin{itemize}
\item[3.] There is at most a single oriented edge with given end points and orientation.
Thus, we can't ``put two edges into the same state'' as a consequence of Fermi statistics.
\end{itemize}
If $R$ has $p$ rows its easy to check that the largest Young diagram that contributes is a block with $p$ columns and $p$ rows.
This corresponds to the Gauss graph with every possible fermion line present.
For example, for $p=3$ we have
\bea
s=\yng(3,3,3)\qquad\longleftrightarrow\qquad\sigma =
\begin{gathered}\includegraphics[scale=0.4]{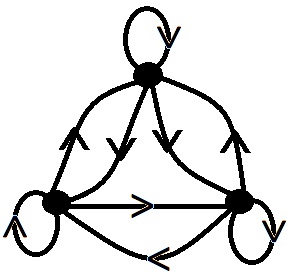}\end{gathered}
\eea
There is often a unique Gauss graph $\sigma$ for each fermionic restricted Schur polynomial, that is the 
restricted Schur polynomial and the Gauss graph bases often coincide.
This is in complete harmony with the results given in \cite{Berenstein:2019esh}, which demonstrate that
in the context of a single fermionic matrix, the Schur polynomial basis and the trace basis are the same.

\section{Asymptotic Symmetries}\label{AsymptoticSymmetries}

In this section we will work out the action of the generators of the $su(2|2)$ global symmetry.
We work in the displaced corners approximation so that impurities located at distinct corners are well separated and
consequently, at large $N$, they are not interacting.
This is the sense in which we mean ``asymptotic'' symmetries.
A nice conclusion of this analysis is that the Gauss graph operators very naturally fall into representations of $su(2|2)$.
Further, we will demonstrate that excitations again carry charges under a central extension of the algebra, generalizing what is
known about the planar limit.

\subsection{Algebra}

The bosonic $su(2)\times su(2)$ subalgebra is generated by $R^a{}_b$ and $L^\alpha{}_\beta$.
The $R^a{}_b$ rotate the bosonic fields $Y,X$ (which are in the $({\bf 2},{\bf 0})$ of the subalgebra) while $L^\alpha{}_\beta$
rotate the fermionic fields $\psi_1,\psi_2$ (which are in the $({\bf 0},{\bf 2})$).
We will refer to these two $su(2)$s as $su(2)_R$ and $su(2)_L$.
In terms of raising and lowering operators
\bea
R^1{}_2=R_+\qquad R^2{}_1=R_-\qquad 2 R^1{}_1=-2R^2{}_2=R_3
\eea
\bea
L^1{}_2=L_+\qquad L^2{}_1=L_-\qquad 2 L^1{}_1=-2L^2{}_2=L_3
\eea
we have
\bea
[R_3,R_-]=-2R_-\qquad [R_3,R_+]=2R_+\qquad [R_+,R_-]=R_3
\eea
and
\bea
[L_3,L_-]=-2L_-\qquad [L_3,L_+]=2L_+\qquad [L_+,L_-]=L_3
\eea
The algebra also has supersymmetry generators $Q^\alpha{}_a$ and $S^a{}_\alpha$.
These generators obey
\bea
[R^a{}_b,Q^\gamma{}_c]=-\delta^a_c Q^\gamma{}_b+{1\over 2}\delta^a_b Q^\gamma{}_c \qquad
[R^a{}_b,S^c{}_\gamma]=\delta^c_b S^a{}_\gamma-{1\over 2}\delta^a_b S^c{}_\gamma
\eea
\bea
[L^\alpha{}_\beta,Q^\gamma{}_c]=\delta^\gamma_\beta Q^\alpha{}_c-{1\over 2}\delta^\alpha_\beta 
Q^\gamma{}_c \qquad
[L^\alpha{}_\beta,S^c{}_\gamma]=-\delta^\alpha_\beta S^\alpha{}_\gamma+{1\over 2}\delta^\alpha_\beta S^c{}_\gamma
\eea
as well as
\bea
\{ Q^\alpha{}_a,S^b{}_\beta\} = \delta^\alpha_\beta R^b{}_a+\delta^b_a L^\alpha{}_\beta +\delta^b_a\delta^\alpha_\beta
C
\eea
\bea
\{ Q^\alpha{}_a,Q^\beta{}_b\} =\epsilon^{\alpha\beta}\epsilon_{ab}P\qquad
\{ S^a{}_\alpha,S^b{}_\beta\} =\epsilon_{\alpha\beta}\epsilon^{ab}K
\label{CentralChargeAlgebra}
\eea
Our goal in the sections that follow is to argue that the state space of the Gauss graph operators are organized into
representations of this algebra, to determine the values of the central charges $P,K$ and $C$ and finally, to demonstrate
that when acting on physical states, the central charges $P$ and $K$ vanish.

\subsection{$SU(2)_R$}

The general state in an $su(2)$ representation can be labeled with a pair of quantum numbers, $j_R,m_R$.
The action of the lowering operator is
\bea
R_-|j_R\, m_R\rangle = \sqrt{j_R(j_R+1)-m_R(m_R-1)}|j_R\, m_R-1\rangle\label{basicRep}
\eea
To determine the representation that a given Gauss graph corresponds to, we identify
\bea
R_-={\rm Tr}\left( X{d\over dY}\right)
\eea
We then act with $R_-$ on a given Gauss graph operator and compare to (\ref{basicRep}).
This analysis is presented in detail in Appendix \ref{RotatingGaussgraphoperators}.
Our conclusion is the following
\begin{itemize}
\item[1.] Each node of the Gauss graph belongs to a definite $SU(2)_R$ representation.
If the number of closed $Y$ loops attached to node $k$ is $b^{(1)}_k$ and the number of closed $X$ loops is $b^{(2)}_k$,
then node $k$ is in the spin $j_R={1\over 2}(b^{(1)}_k+b^{(2)}_k)$ representation.
\item[2.] The specific state in the representation that node $k$ occupies is determined by 
$m_R={1\over 2}(b^{(1)}_k-b^{(2)}_k)$. 
\item[3.] The action of $R_-$ on the $k$th node replaces a single directed $Y$ edge with a single directed $X$ edge, with an
overall coefficient given by (\ref{basicRep}).
\item[4.] The generators $R^a{}_b$ do not act on edges that travel between nodes.
\end{itemize}
From the action defined for $R_-$ above we can work out the action of $R_+$ (by hermittian conjugation) and the
action of $R_3$ (by using the $su(2)_R$ algebra).

The complete action of the $su(2)_R$ generators follows by summing the result of acting on each node in the graph.
This corresponds to the usual co-product action.
Notice that in moving to the Gauss graph basis, we have in fact organized the state space into $su(2)_R$ multiplets.

\subsection{$SU(2)_L$}

In this case we identify
\bea
L_-={\rm Tr}\left(\psi_2{d\over d\psi_1}\right)
\eea
We again find that the $su(2)$ generators (now the $L^\alpha{}_\beta$) do not act on edges that travel between nodes.
Each node is again in a definite state. We find four possibilities
\begin{itemize}
\item[1.] A node that has no closed $\psi_1$ loops and no closed $\psi_2$ loops is in the one dimensional representation
with $j_L=0$.
\item[2.] A node with a single closed $\psi_1$ loop is in the representation $j_L={1\over 2}$, and in state $m_L={1\over 2}$.
$L_-$ acting on this node replaces the $\psi_1$ loop with a $\psi_2$ loop and $L_+$ annihilates the node. 
\item[3.] A node with a single closed $\psi_2$ loop is in the representation $j_L={1\over 2}$, and in state $m_L=-{1\over 2}$.
$L_-$ annihilates the node while $L_+$ acting on this node replaces the $\psi_2$ loop with a $\psi_1$ loop.
\item[4.] A node that has both a closed $\psi_1$ loop and a closed $\psi_2$ loops is in the one dimensional representation
with $j_L=0$.
\end{itemize}
As in the previous section, the complete action of the $su(2)_L$ generators follows by summing the result of acting on 
each node in the graph.
Further, as above, the Gauss graph basis is organized into $su(2)_L$ multiplets.

\subsection{Supercharges}

When the supercharges act we will again assume that there is an action on each node of the graph and that the total
action is the sum of actions on each node.
In what follows it is more convenient to specify the Gauss graph by stating how many closed loops of each species there are
at each node and how many edges (with orientation) there are stretching between nodes.
The numbers $b^{(a)}_k$ count the number of closed bosonic edges at node $k$, while $f^{(\alpha)}_k$ count the number
of closed fermionic edges at node $k$.
The numbers  $b^{(a)}_{ij}$ count the number of bosonic edges moving from node $i$ to node $j$, while
$f^{(\alpha)}_{ij}$ count the number of fermionic edges moving from node $i$ to node $j$.
We will assume the following action for the supercharges, acting on node $i$
\bea
&&(Q^\alpha{}_a)_i O_{R,r}(\{\cdots,b^{(c)}_i,f^{(\gamma)}_i,\cdots\})
=c_a(1-f^{(\alpha)}_i)\sqrt{b^{(a)}_i}\,
O_{R,r}(\{\cdots,b^{(c)}_i-\delta^c_a,f^{(\gamma)}_i+\delta^\gamma_\alpha,\cdots\})\cr
&&\qquad +c_b\sum_{b=1}^2\sum_{\beta=1}^2 f^{(\beta)}_i\epsilon^{\alpha\beta}\epsilon_{ab}
\sqrt{b^{(b)}_i+1}\,O_{R_i^+,r_i^+}(\{\cdots,b^{(c)}_i+\delta^c_b,f^{(\gamma)}_i-\delta^\gamma_\beta,\cdots\})
\eea
\bea
&&(S^a{}_\alpha)_i O_{R,r}(\{\cdots,b^{(c)}_i,f^{(\gamma)}_i,\cdots\})
=c_d\, f^{(\alpha)}_i\sqrt{b^{(a)}_i+1}\,
O_{R,r}(\{\cdots,b^{(c)}_i+\delta^c_a,f^{(\gamma)}_i-\delta^\gamma_\alpha,\cdots\})\cr
&&\qquad +c_c\sum_{b=1}^2\sum_{\beta=1}^2 (1-f^{(\beta)}_i)\epsilon_{\alpha\beta}\epsilon^{ab}
\sqrt{b^{(b)}_i}\,O_{R_i^-,r_i^-}(\{\cdots,b^{(c)}_i-\delta^c_b,f^{(\gamma)}_i+\delta^\gamma_\beta,\cdots\})
\eea
In the argument of $O_{R,r}$ we have only explicitly specified quantum numbers of the state that change under the action of 
the supercharge. 
Notice that both supercharges change the shape of the Young diagram labels $R$ and $r$; see 
Appendix \ref{YoungDiagramNotations} for an explanation of this notation.
The two labels $R$ and $r$ change in precisely the same way.
The coefficients $c_a,c_b,c_c$ and $c_d$ are constants that will be determined by requiring that $Q^\alpha{}_a$ and
$S^a{}_\alpha$ close the correct algebra.
The factor of $f^{(\alpha)}_i$ and $(1-f^{(\alpha)}_i)$ are there to ensure that we don't put two fermions into one state
or remove a fermion from a state that doesn't contain any.
The factors of $\sqrt{b^{(a)}_i}$ and $\sqrt{b^{(a)}_i+1}$ are there for convenience.
With these factors, the coefficients $c_a,c_b,c_c$ and $c_d$ are independent of $b^{(a)}_i$.
The factors of $\epsilon^{ab}$ and $\epsilon^{\alpha\beta}$ are determined by $su(2)_R\times su(2)_L$ covariance.

The above ansatz is strongly motivated by the action of the supercharges worked out in \cite{Beisert:2005tm}.
The key differences are
\begin{itemize}
\item[1.] The excitations of \cite{Beisert:2005tm} are either a single $Y$ or a single $X$ field. 
Here we can have an arbitrary number of both.
The only effect is that we now need to include the $\sqrt{b^{(a)}_i}$ and $\sqrt{b^{(a)}_i+1}$ factors.
\item[2.] The fermionic states can have any occupancy. This is why we need the $f^{(\alpha)}_i$ and $(1-f^{(\alpha)}_i)$
factors.
\item[3.] The action of \cite{Beisert:2005tm} was written down using markers ${\cal Z}^\pm$, which insert or remove
$Z$s from the single trace operator, leading to a dynamic lattice with a time dependent number of sites.
Here we have a truly non-planar generalization of this action: a box is added or deleted to the Young diagram labels.
It appears to be highly non-trivial to describe this operation in terms of traces.
\end{itemize}

Our next task is to show that these supercharges close the correct algebra and, in the process determine the coefficients
$c_a,c_b,c_c$ and $c_d$, as well as the values of the central extensions.

\subsection{Representation}

To begin we require that
\bea
\{ (Q^\alpha{}_a)_i ,(S^b{}_\beta)_i\} = \delta^\alpha_\beta (R^b{}_a)_i+\delta^b_a (L^\alpha{}_\beta)_i 
+\delta^b_a\delta^\alpha_\beta C_i
\eea
This forces
\bea
  c_a c_d-c_b c_c=1
\eea
and the central charge is
\bea
  C_i = {1\over 2}(b^{(1)}+b^{(2)}+f^{(1)}+f^{(2)})
\eea
The central extension vanishes
\bea
\{ (Q^\alpha{}_a)_i ,(Q^\beta{}_b)_i\} =0
\eea
\bea
\{ (S^a{}_\alpha)_i ,(S^b{}_\beta)_i\} =0
\eea
This is the correct description of the free theory.
In particular, we find that there are no anomalous dimensions.
This is not correct when interactions are turned on: the Gauss graph operators are not in general BPS and they will 
develop non-zero anomalous dimensions.
Indeed, looking at the one loop result (\ref{lovelyanswer}) it is clear that this is the case.
Studying (\ref{lovelyanswer}) leads to a second puzzle: at least at one loop, the anomalous dimension depends only on
$n_{ij}=b^{(1)}_{ij}+b^{(2)}_{ij}+f^{(1)}_{ij}+f^{(2)}_{ij}$.
These are quantum numbers associated to edges that stretch between different nodes.
This dependence appears puzzling because our analysis thus far has demonstrated that the global symmetry generators leave
these edges inert!

It is not hard to appreciate why the global symmetry generators do not act on these edges.
An edge forming a closed loop at a node is automatically gauge invariant.
In contrast to this, edges going between nodes are constrained by the requirement of gauge invariance to form closed
paths that respect the orientation of each edge.
Replacing one edge with another edge of a different species spoils the Gauss law constraint so that we land up with
a state that is not gauge invariant. 
Thus, the edges that straddle nodes are not transformed by the global symmetry generators because there is no gauge 
invariant state that they could be transformed into.
If however we act with a pair of supercharges (for example) we can change the species of an edge with the first action 
and restore it with the second.
Consequently, the edges straddling nodes can give rise to the central extensions introduced below
\bea
\{ (Q^\alpha{}_a)_i ,(Q^\beta{}_b)_j\} =\epsilon^{\alpha\beta}\epsilon_{ab}P_{ij}
\qquad
\{ (S^a{}_\alpha)_i ,(S^b{}_\beta)_j\} =\epsilon_{\alpha\beta}\epsilon^{ab}K_{ij}
\eea
Our proposal for the action of the central extensions on an excitation stretching between the nodes of a Gauss graph operator is
\bea
P_{ij}O_{R,r}(\sigma)=\alpha\sqrt{N+r_i}O_{R_i^+,r_i^+}(\sigma)
-\alpha\sqrt{N+r_j}O_{R_j^+,r_j^+}(\sigma)
\eea
\bea
K_{ij}O_{R,r}(\sigma)=\beta\sqrt{N+r_i}O_{R_i^-,r_i^-}(\sigma)
-\beta\sqrt{N+r_j}O_{R_j^-,r_j^-}(\sigma)
\eea
These formulas are the natural generalization of the action of the central extension obtained in \cite{Beisert:2005tm}.
Indeed, the markers ${\cal Z}^\pm$ are again replaced by an action that adds or removes a box from the Young diagram.
Further, these actions again reveal the nature of the central extension as a gauge transformation, exactly as was observed
in the planar limit.
An important consistency condition is that these central extensions must vanish when acting on physical states.
In the planar limit this follows from cyclicity of the trace.
In the non-planar problem we study here we find that
\bea
\sum_{i,j}P_{ij}=0=\sum_{i,j}K_{ij}
\eea
holds as a consequence of the Gauss Law constraint.
The fact that the Gauss graph operators are gauge invariant physical states implies that they are annihilated by the total 
central extension.

Using the above central extension we obtain the following formula for the anomalous dimension $\gamma$ of a Gauss graph
operator
\bea
\gamma={1\over 2}\sum_{ij}\sqrt{1+P_{ij}K_{ij}}
\eea
To see that this correctly reproduces the one loop anomalous dimension, note that
\bea
P_{ij}K_{ij}O_{R,r}(\sigma)&=&\alpha\beta \Big[(N+r_i)O_{R,r}(\sigma)+(N+r_j)O_{R,r}(\sigma)\cr\cr
&&-\,\, \sqrt{(N+r_i)(N+r_j)}\left(O_{R_{ij}^+,r_{ij}^-}(\sigma)+O_{R_{ij}^-,r_{ij}^-}(\sigma)\right)\Big]\cr\cr
&&
\eea
which, after summing over $i$ and $j$ and setting $\alpha\beta=g_{YM}^2$ is nothing but (\ref{lovelyanswer}).

\section{Discussion}\label{Discussion}

Our main result is the decomposition of the state space of CFT operators dual to excited giant graviton branes into
irreducible representations of the $su(2|2)\ltimes\BR$ global symmetry.
There are a number of positive features of our results which support their validity:
\begin{itemize}
\item[1.] Our analysis shows that the state space of restricted Schur polynomials is not organized into irreducible 
representations of the $su(2|2)\ltimes\BR$ global symmetry. 
However, after transforming to the Gauss graph operator basis, we do indeed have a transparent $su(2|2)\ltimes\BR$
structure. Indeed, it is a simple matter to read off the $su(2)_R\times su(2)_L$ quantum numbers from the graph.
\item[2.] We have managed to reproduce the one loop anomalous dimension of the Gauss graph operator from the 
$su(2|2)\ltimes\BR$ central charge. This central charge makes a prediction for the higher loop anomalous dimensions.
It would be interesting to check these predictions. 
\item[3.] Further, excitations are again charged under a central extension of global symmetry. 
Since the original global symmetry is not centrally extended, the action of the central extension must vanish on physical states.
In planar the limit the central extension generates gauge transformations and hence the central extension vanishes when
acting on physical states which are gauge invariant.
In our case the central charge is again set to zero by gauge invariance: the constraint enforced by the Gauss Law ensures
that the central extension vanishes.
Further, the central extension again generates gauge transformations. 
\end{itemize}
This is compelling evidence in support of our results.

There are a number of directions in which our study can be extended.
One could for example try to formulate a more complete description of excited gaint graviton states, by relaxing the restriction
to the $su(2|3)$ sector.
In this case the global symmetry algebra is $su(2|2)\times su(2|2)\ltimes \BR$.
This has proved to be a very fruitful direction in the planar limit of the theory. 
Another fascinating direction would be to use the global symmetry to study interactions of the excitations.
Following \cite{Beisert:2005tm}, a productive way forwards maybe to introduce an $S$-matrix and to use the global
symmetry to constrain its form.
The Gauss graph operators are natural asymptotic states that might be used to define an $S$-matrix.
For example, consider the following (schematic) state
\bea
|{\rm in}\rangle =\young(\,\,\,\,\,\,\,\,\,\,AA,\,\,\,\,\,BB)\cr
\eea
which we will treat as an ``in state''.
Under time evolution by the dilatation operator, the lengths of the rows can change.
When the row lengths are comparable the two impurities can interact, and possibly even swap the row they belong to
or rearrange in even more complicated ways.
The rows lengths will then continue to evolve until the impurities are again well separated, defining an ``out state'' of the
schematic form
\bea
|{\rm out}\rangle =\young(\,\,\,\,\,\,\,\,\,\,BB,\,\,\,\,\,AA)\cr
\eea
The map from the in state to the out state
\bea
|{\rm out}\rangle = S|{\rm in}\rangle
\eea
defines an $S$-matrix as usual.
In the planar case there is a lot one can do with the $S$-matrix.
The powerful methods of integrability can be applied thanks to the fact that the $S$-matrix satisfies a Yang-Baxter equation,
which expresses the equality of two particle scattering between three particles, with the two particle scattering taking place 
in different orders.
Here there is a natural analog of this setup: consider a Young diagram $R$ with three rows, and a Gauss graph operator
that has excitations on each row. 
One can ask if there is equality between the different orders in which the excitations on the different rows can scatter.
Do we obtain something like the Yang-Baxter equation? 
Is it possible to generalize something of the powerful integrability machinery? 
This is the subject of work in progress.

\medskip 

\begin{center} 
{ \bf Acknowledgements}
\end{center} 

We thank Sanjaye Ramgoolam for helpful discussions. 
This work is supported by the South African Research Chairs
Initiative of the Department of Science and Technology and National Research
Foundation as well as funds received from the National Institute for
Theoretical Physics (NITheP).

\vskip0.2cm

\appendix

\section{Young Diagram Notations}\label{YoungDiagramNotations}

The dilatation operator $D$, central charges $C$, $P_{ij}$ and $K_{ij}$ as well as the supercharges $Q^\alpha{}_a$ and
$S^a{}_\alpha$, when acting on the Gauss graph operator $O_{R,r}(\sigma)$, have a non-trivial action on the Young 
diagram labels $R$ and $r$.
In this Appendix we will briefly spell out the notation we use, with a few examples to illustrate the ideas.
Consider the Young diagram $r$ given by
\bea
r=\yng(10,6,4)
\eea
The dilatation operator can transport a box from row $i$ to row $j$.
We use the notation $r_{ij}^+$ to describe the Young diagram obtained from $r$ by deleting a box from row $j$
and adding a box to row $i$.
As an example, we give
\bea
r_{12}^+=\yng(11,5,4)
\eea
We will also find it convenient to use the notation $r_{ij}^-$ to describe the Young diagram obtained from $r$ by deleting 
a box from row $i$ and adding a box to row $j$.
As an example of this notation, consider
\bea
r_{12}^-=\yng(9,7,4)
\eea
Notice that $r_{ij}$, $r_{ij}^+$ and $r_{ij}^-$ all have the same number of boxes.
The supercharges change the number of boxes in the Young diagram.
For example, $Q^\alpha{}_a$ can add a box to a given row.
We use $r_i^+$ to denote the Young diagram obtained from $r$ by adding a single box to row $i$.
For example
\bea
r_{2}^+=\yng(10,7,4)
\eea
Notice the the number of boxes is not preserved: $r_{2}^+$ has one more box that $r$.
The supercharge $S^a{}_\alpha$ can remove a box from a given row.
We use $r_i^-$ to denote the Young diagram obtained from $r$ by deleting a single box from row $i$.
As an example of this notation, we quote
\bea
r_{2}^-=\yng(10,5,4)
\eea
Finally, although we have illustrated the notation using Young diagram $r$, the discussion also holds for $R$.

\section{Restricted Schur Polynomials with 2 rows}\label{RestrictedSchurPolynomialswith2rows}

A simple setting in which to test the formulas and ideas developed in this study, is to consider Young diagrams $R$
that have two rows.
The problem with two rows (or columns) is particularly simple because upon restricting an irreducible representation of $S_n$ 
to any subgroup $S_k\times S_{n-k}$, irreducible representations of the subgroup appear without multiplicity. 
In Appendix \ref{RotatingRestrictedSchurPolynomials} we evaluate the action of $su(2)$ rotations on restricted
Schur polynomials with bosonic excitations only. 
Since there are no mulitplicities, the relevant restricted Schur polynomials are $\chi_{R,(b_0,b_1,b_2)}(Z,Y,X)$.
There is a $S_{b^{(1)}}\times S_{b^{(2)}}$ symmetry that is Schur Weyl dual to $U(2)$. 
Consequently, the projection operators needed to construct the restricted Schur polynomials are easily determined in 
terms of well known $SU(2)$ Clebsch-Gordan coefficients\cite{Carlson:2011hy}.
We use the quantum numbers $j,j_3$ for the $SU(2)$ used to organize the $Y$ fields and $k,k_3$ for the $SU(2)$ used 
to organize the $X$ fields.

Let $(b_i)_k$ denote the number of boxes in row $k$ of Young diagram $b_i$.
The translation of the restricted Schur polynomial $\chi_{R,(b_0,b_1,b_2)}(Z,Y,X)$ to $SU(2)$ state labels is as follows
\bea
\begin{array}{ccc}
   (b_2)_1={p\over 2}+k  &\qquad &(b_2)_2={p\over 2}-k\\
\\
   (b_1)_1={m\over 2}+j &\qquad &(b_1)_2={m\over 2}-j\\
\\
   R_1=(b_0)_1+{m+p\over 2}+j_3+k_3 &\qquad &R_2=(b_0)_2+{m+p\over 2}-j_3-k_3
\end{array}
\label{translate}
\eea
$j_3$ is equal to the number of $Y$ boxes in the first row of $R$ minus the number of $Y$ boxes in the second.
$k_3$ is defined in the same way, but for the $X$ boxes. 
The above labels may appear to be over complete: given $b^{(0)},b^{(1)},b^{(2)}$ as well as $b_0,k,j,k_3+j_3$ we can reconstruct the Young diagram labels $R,b_0,b_1$ and $b_2$.
It seems that we need only the sum $k_3+j_3$ and not the individual values $j_3,k_3$.
The point is that, even when $R$ has two rows, when we restrict $S_{a+b+c}$ to $S_a\times S_b\times S_c$
we do need a multiplicity label.
Specifying $k_3$ and $j_3$ independently resolves the multiplicity - its tells us which boxes in $R$ are $Y$ boxes and
which are $X$ boxes.
The simplest way to see this is to note that we can first restrict $S_{b^{(0)}+b^{(1)}+b^{(2)}}$ to 
$S_{b^{(2)}}\times S_{b^{(0)}+b^{(1)}}$ without multiplicity, and then restrict $S_{b^{(0)}+b^{(1)}}$ 
to $S_{b^{(1)}}\times S_{b^{(0)}}$, again without multiplicity.
The first restriction introduces $(k,k_3)$ and the second $(j,j_3)$.

\section{Rotating Restricted Schur Polynomials}\label{RotatingRestrictedSchurPolynomials}

In this Appendix we review results that were obtained in \cite{Bornman:2016zwn}.
We would like to obtain the action of the following $su(2)_R$ generators
\bea
  &&R_-={\rm Tr}\left( X{d\over dY}\right)\qquad
  R_+={\rm Tr}\left( Y{d\over dX}\right)\cr
  &&R_3=[R_+,R_-]={\rm Tr}\left(Y{d\over dY}-X{d\over dX}\right)
\eea
Once we have evaluated the action of $R_+$, the action of $R_-$ follows by hermittian conjugation, and the
action of $R_3$ then follows by using the $su(2)$ algebra.
Consequently, we only need the action of $R_-={\rm Tr}\left( X{d\over dY}\right)$.
The computation is carried out by allowing $R_-$ to act on the restricted Schur polynomial.
The result can then be expressed as a linear combination of restricted Schur polynomials, since the restricted Schur
operators provide a basis.
The coefficients of this linear expansion are given by the trace of a product of projection operators.
In the distant corners approximation, the computation of the traces that need to be computed is reduced to 
the evaluation of $su(2)$ Clebsch-Gordan coefficients.
The result is \cite{Bornman:2016zwn}
\bea
{\rm Tr}\left( X{d\over dY}\right)O^{(n,m,p)}_{R,r,j,j_3,k,k_3}\cr
={j+j_3\over 2j}{k+k_3+1\over 2k+1}
\sqrt{\Big({m\over 2}+j+1\Big){2j\over 2j+1}}
\sqrt{\Big({p\over 2}+k+2\Big){2k+1\over 2k+2}}
O^{(n,m-1,p+1)}_{R,r,j-{1\over 2},j_3-{1\over 2},k+{1\over 2},k_3+{1\over 2}}\cr
+{j+j_3\over 2j}{k-k_3\over 2k+1}
\sqrt{\Big({m\over 2}+j+1\Big){2j\over 2j+1}}
\sqrt{\Big({p\over 2}-k+1\Big){2k+1\over 2k}}
O^{(n,m-1,p+1)}_{R,r,j-{1\over 2},j_3-{1\over 2},k-{1\over 2},k_3+{1\over 2}}\cr
+{j-j_3+1\over 2j+2}{k+k_3+1\over 2k+1}
\sqrt{\Big({m\over 2}-j\Big){2j+2\over 2j+1}}
\sqrt{\Big({p\over 2}+k+2\Big){2k+1\over 2k+2}}
O^{(n,m-1,p+1)}_{R,r,j+{1\over 2},j_3-{1\over 2},k+{1\over 2},k_3+{1\over 2}}\cr
+{j-j_3+1\over 2j+2}{k-k_3\over 2k+1}
\sqrt{\Big({m\over 2}-j\Big){2j+2\over 2j+1}}
\sqrt{\Big({p\over 2}-k+1\Big){2k+1\over 2k}}
O^{(n,m-1,p+1)}_{R,r,j+{1\over 2},j_3-{1\over 2},k-{1\over 2},k_3+{1\over 2}}\cr
+{j-j_3\over 2j}{k-k_3+1\over 2k+1}
\sqrt{\Big({m\over 2}+j+1\Big){2j\over 2j+1}}
\sqrt{\Big({p\over 2}+k+2\Big){2k+1\over 2k+2}}
O^{(n,m-1,p+1)}_{R,r,j-{1\over 2},j_3+{1\over 2},k+{1\over 2},k_3-{1\over 2}}\cr
+{j-j_3\over 2j}{k+k_3\over 2k+1}
\sqrt{\Big({m\over 2}+j+1\Big){2j\over 2j+1}}
\sqrt{\Big({p\over 2}-k+1\Big){2k+1\over 2k}}
O^{(n,m-1,p+1)}_{R,r,j-{1\over 2},j_3+{1\over 2},k-{1\over 2},k_3-{1\over 2}}\cr
+{j+j_3+1\over 2j+2}{k-k_3+1\over 2k+1}
\sqrt{\Big({m\over 2}-j\Big){2j+2\over 2j+1}}
\sqrt{\Big({p\over 2}+k+2\Big){2k+1\over 2k+2}}
O^{(n,m-1,p+1)}_{R,r,j+{1\over 2},j_3+{1\over 2},k+{1\over 2},k_3-{1\over 2}}\cr
+{j+j_3+1\over 2j+2}{k+k_3\over 2k+1}
\sqrt{\Big({m\over 2}-j\Big){2j+2\over 2j+1}}
\sqrt{\Big({p\over 2}-k+1\Big){2k+1\over 2k}}
O^{(n,m-1,p+1)}_{R,r,j+{1\over 2},j_3+{1\over 2},k-{1\over 2},k_3-{1\over 2}}\, .\cr
\eea
These are not exact expressions - there are corrections of order ${b^{(1)}\over b^{(0)}}$ and ${b^{(2)}\over b^{(0)}}$,
which are subleading at large $N$.
Notice that there is a complicated mixing of the restricted Schur polynomials under $su(2)_R$.
The restricted Schur polynomials are not organized into multiplets of $su(2)_R$

\section{Gauss Graph Transformations}\label{GaussGraphTransformations}

In this Appendix we will derive explicit formulas for the transformation from the restricted Schur polynomial basis
to the Gauss graph basis. 
These transformation formulas are needed to
\begin{itemize}
\item[1.] Construct the Hilbert space of the excited giant graviton brane system.
\item[2.] Translate the action of $su(2)$ generators from the restricted Schur basis to the Gauss graph basis.
\end{itemize}
 
\subsection{Bosonic Operators}

As a non-trivial example of how we move from the restricted Schur basis to the Gauss graph basis, consider an excitation
constructed using 4 bosonic $Y$ fields.
Assume that we study a 2 brane system so that both $R$ and $r$ have two rows.
We remove two excitations from each row so that
\bea
R={\tiny\yng(10,6)}\qquad r={\tiny\yng(10,6)}
\eea
Denoting the excitations removed from row 1 by $1,2$ and the excitations removed from row 2 by $3,4$ we have
\bea
H=\{1,(12),(34),(12)(34)\}
\eea
In the restricted Schur basis, the possible representation that the excitations can be arranged into are
\bea
s\in\left\{{\tiny\yng(4)}\, ,\,{\tiny \yng(3,1)}\, ,\,{\tiny \yng(2,2)}\right\}
\eea
We choose our permutation so that we are describing a pair of strings stretched between nodes 1 and 2
\bea
\sigma=(13)(24)=\begin{gathered}\includegraphics[scale=0.4]{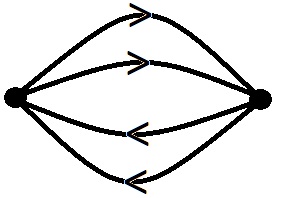}\end{gathered}
\eea
We would like to compute the transformation coefficients, given by
\bea
C^{(s)}\left((13)(24)\right)={|H|\over\sqrt{b^{(1)}}!}\sqrt{d_s}\sum_{k,m=1}^{d_s}
\Gamma^{(s)}\left((13)(24)\right)_{km}B_k^{s\to 1_H} B_m^{s\to 1_H}
\eea
There are no multiplicity labels on the branching coefficient because $R$ has 2 rows.
The branching coefficient is determined by
\bea
   {1\over |H|}\sum_{\gamma\in H}\Gamma^{(s)}(\gamma)_{km}=B_k B_m
\eea
For $s={\tiny \yng(4)}$ the representation is one dimensional, $\Gamma^{({\tiny\yng(4)})}(\sigma)=1$ for any $\sigma$ and
the branching coefficient $B=1$.
Consequently
\bea
C^{\left({\tiny\yng(4)}\right)}\left( (13)(24)\right)={4\over\sqrt{24}}\cdot\sqrt{1}\cdot 1=\sqrt{2\over 3}
\eea
For $s={\tiny \yng(3,1)}$ the representation $\Gamma^{\left({\tiny\yng(3,1)}\right)}(\sigma)$ is three dimensional.
The branching coefficient is determined to be
\bea
B=\left[\begin{array}{c} {1\over\sqrt{3}}\\\sqrt{2\over 3}\\ 0\end{array}\right]
\eea
and consequently
\bea
C^{\left({\tiny\yng(3,1)}\right)}\left( (13)(24)\right)={4\over\sqrt{24}}\cdot\sqrt{3}\cdot 
\Gamma^{\left({\tiny \yng(3,1)}\right)}_{km}B_k B_m =-\sqrt{2}
\eea
Finally for $s={\tiny \yng(2,2)}$ the representation $\Gamma^{\left({\tiny\yng(2,2)}\right)}(\sigma)$ is two dimensional.
The branching coefficient is determined to be
\bea
B=\left[\begin{array}{c} {0}\\ 1\end{array}\right]
\eea
and consequently
\bea
C^{\left({\tiny\yng(3,1)}\right)}\left( (13)(24)\right)={4\over\sqrt{24}}\cdot\sqrt{2}\cdot 
\Gamma^{\left({\tiny \yng(2,2)}\right)}_{km}B_k B_m ={2\over \sqrt{3}}
\eea
Thus, we find that
\bea
O_{R,r}\left(\begin{gathered}\includegraphics[scale=0.3]{graphss_4}\end{gathered}\right)=
\sqrt{2\over 3}O_{R,(r,{\tiny \yng(4)})}-\sqrt{2}O_{R,(r,{\tiny \yng(3,1)})}
+{2\over\sqrt{3}}O_{R,(r,{\tiny \yng(2,2)})}
\eea
We did not explicitly specify that we remove two impurities from the first row and two from the second row on the
right hand side of this equation, but it can be read off of the graph appearing on the left hand side.

Here are a few more examples of transformations between the restricted Schur and Gauss graph bases
\bea
O_{R,r}\left(\begin{gathered}\includegraphics[scale=0.4]{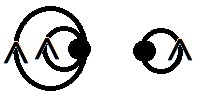}\end{gathered}\right)=
\sqrt{2\over 3}O_{R,(r,{\tiny \yng(3)})}+{2\over\sqrt{3}}O_{R,(r,{\tiny \yng(2,1)})}
\eea
\bea
O_{R,r}\left(\begin{gathered}\includegraphics[scale=0.4]{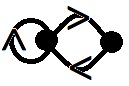}\end{gathered}\right)=
\sqrt{2\over 3}O_{R,(r,{\tiny \yng(3)})}-{1\over\sqrt{3}}O_{R,(r,{\tiny \yng(2,1)})}
\eea
\bea
O_{R,r}\left(\begin{gathered}\includegraphics[scale=0.4]{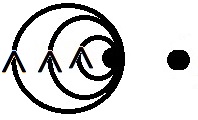}\end{gathered}\right)=\sqrt{6}
O_{R,(r,{\tiny \yng(3)})}
\eea
The last example above generalizes very nicely: for $m$ loops attached to the first node, we replace $s$ by a
Young diagram that is a single row with $m$ boxes.
These expression will be very useful in Appendix \ref{RotatingGaussgraphoperators} when we study the action of
rotations on Gauss graph operators, using the known action of rotations on restricted Schur polynomials.

\subsection{Fermionic Operators}

The structure of the state space of the fermionic Gauss graphs depends on properties of the transformation
from restricted Schur polynomials to Gauss graph operators.
For that reason we work out a few carefully chosen examples in this Appendix.
Consider an excitation constructed from the $\psi_1$ field.
The transformation coefficients from the representation $s$ and multiplicity labels $\mu_1,\mu_2$ that 
organize the fermionic excitations, to permutation $\tau$ are given by
\bea
\tilde C^{(s)}_{\mu_1\mu_2}(\tau)=|H|\sqrt{d_s\over f^{(1)}!}
\sum_{k,m=1}^{d_s}\left(\Gamma^{(s)}(\tau)\hat{O}\right)_{km}
B^{s\to 1_H}_{k\mu_1}B^{s^T\to 1^{f^{(1)}}}_{m\mu_2}
\eea
We have used $d_s=d_{s^T}$.
Notice that two distinct branching coefficients appear.
Before evaluating any examples of the coefficients $\tilde C^{(s)}_{\mu_1\mu_2}(\tau)$ we will relate the two 
branching coefficients that appear.
Starting from the definition of the branching coefficient $B^{s^T\to 1^{f^{(1)}}}_{m\mu}$ we easily find
\bea
\sum_\mu B^{s^T\to 1^{f^{(1)}}}_{k\mu}B^{s^T\to 1^{f^{(1)}}}_{m\mu}
&=&{1\over |H|}\sum_{\gamma\in H}{\rm sgn}(\gamma)\Gamma^{(s^T)}(\gamma)_{km}\cr
&=&{1\over |H|}\sum_{\gamma\in H}{\rm sgn}(\gamma)(\hat{O}\Gamma^{(s)}(\gamma)\hat{O})_{km}\cr
&=&{1\over |H|}\sum_{\gamma\in H}\Gamma^{(s)}(\gamma)_{km}\cr
&=&\sum_\mu B^{s\to 1_H}_{k\mu}B^{s\to 1_H}_{m\mu}
\eea
which proves that the two branching coefficients are in fact equal!
Consequently the formula for the transformation coefficients can be simplified to
\bea
\tilde C^{(s)}_{\mu_1\mu_2}(\tau)=|H|\sqrt{d_s\over f^{(1)}!}
\sum_{k,m=1}^{d_s}\left(\Gamma^{(s)}(\tau)\hat{O}\right)_{km}
B^{s\to 1_H}_{k\mu_1}B^{s\to 1_H}_{m\mu_2}
\eea
In what follows we again restrict to examples for which $R$ has two rows so that we can again drop multiplicity labels.

To begin, consider an excitation constructed using three $\psi_1$s.
Two of the $\psi_1$ impurities live in the first row of $R$ and one in the second row.
The only possible representation that leads to a non-zero restricted Schur polynomial is $s={\tiny \yng(2,1)}$
as already explained in Section \ref{FermionicStateSpace}.
A simple computation shows that
\bea
\hat{O}=\left[\begin{array}{cc} 0 & 1\\ -1 & 0 \end{array}\right]
\eea
The group $H=\{1, (12)\}$ and the branching coefficient is
\bea
B=\left[\begin{array}{c}{\sqrt{3}\over 2} \\ {1\over 2}\end{array}\right]
\eea
It is now straight forward to verify that
\bea
\tilde{C}^{\left({\tiny\yng(2,1)}\right)}\left(\begin{gathered}\includegraphics[scale=0.4]{graphss_5}\end{gathered}\right)=
\tilde{C}^{\left({\tiny\yng(2,1)}\right)}\left(1\right)=\tilde{C}^{\left({\tiny\yng(2,1)}\right)}\left(\,(12)\,\right)=0
\eea

\bea
\tilde{C}^{\left({\tiny\yng(2,1)}\right)}\left(\begin{gathered}\includegraphics[scale=0.4]{graphss_6}\end{gathered}\right)
&=&\tilde{C}^{\left({\tiny\yng(2,1)}\right)}\left(\,(13)\,\right)=-\tilde{C}^{\left({\tiny\yng(2,1)}\right)}\left(\,(23)\,\right)\cr
&=&\tilde{C}^{\left({\tiny\yng(2,1)}\right)}\left(\,(132)\,\right)
=-\tilde{C}^{\left({\tiny\yng(2,1)}\right)}\left(\,(123)\,\right)=1
\eea
The negative signs which appear above are exactly what we expect.
They reflect an odd number of swaps of fermion fields.

For the second example, consider an excitation constructed using four $\psi_1$s and again consider a Young diagram 
$R$ with two rows.
Two of the $\psi_1$ impurities live in the first row of $R$ and two in the second row.
The only possible representation that leads to a non-zero restricted Schur polynomial is $s={\tiny \yng(2,2)}$,
which was also explained in Section \ref{FermionicStateSpace}.
A straight forward computation shows that we again have
\bea
\hat{O}=\left[\begin{array}{cc} 0 & 1\\ -1 & 0 \end{array}\right]
\eea
The group $H=\{1, (12), (34), (12)(34)\}$ and the branching coefficient is easily determined to be
\bea
B=\left[\begin{array}{c}0 \\ 1\end{array}\right]
\eea
It is now straight forward to verify that
\bea
\tilde{C}^{\left({\tiny\yng(2,2)}\right)}\left(\begin{gathered}\includegraphics[scale=0.4]{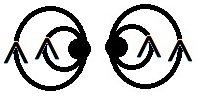}\end{gathered}\right)=0
\eea
\bea
\tilde{C}^{\left({\tiny\yng(2,2)}\right)}\left(\begin{gathered}\includegraphics[scale=0.4]{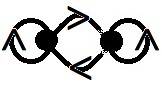}\end{gathered}\right)=0
\eea
\bea
\tilde{C}^{\left({\tiny\yng(2,2)}\right)}\left(\begin{gathered}\includegraphics[scale=0.35]{graphss_4}\end{gathered}\right)=1
\eea

\section{Rotating Gauss graph operators}\label{RotatingGaussgraphoperators}

In this section we will use the action of the $su(2)_R$ generators on restricted Schur polynomials given in 
Appendix \ref{RotatingRestrictedSchurPolynomials}, and the translation between restricted Schur polynomials and Gauss
graphs worked out in Appendix \ref{GaussGraphTransformations}, to determine the action of the $su(2)_R$ generators on
the Gauss graph operators.

To begin we will work out an example which demonstrates that the $su(2)_R$ generators leave the edges in a Gauss graph, that
stretch between distinct nodes, inert.
The computation is most easily phrased using the notation introduced in Appendix \ref{RestrictedSchurPolynomialswith2rows}.
Consider a two giant system constructed using $b^{(0)}$ $Z$ fields, 4 $Y$ fields and no $X,\psi_1$ or $\psi_2$ fields.
Two $Y$ fields belong to the first row of $R$ and two to the second row.
Our starting point is the formula
\bea
O_{R,r}\left(\begin{gathered}\includegraphics[scale=0.35]{graphss_4}\end{gathered}\right)=
\sqrt{2\over 3}O_{R,r,2,0,0,0}-\sqrt{2}O_{R,r,1,0,0,0}+{2\over\sqrt{3}}O_{R,r,0,0,0,0}
\eea
A simple application of the formula in Appendix \ref{RotatingRestrictedSchurPolynomials} leads to
\bea
{\rm Tr}\left(X{d\over dY}\right) O_{R,r,2,0,0,0}&=&
O_{R,r,{3\over 2},-{1\over 2},{1\over 2},{1\over 2}}+
O_{R,r,{3\over 2},{1\over 2},{1\over 2},-{1\over 2}}
\eea
\bea
{\rm Tr}\left(X{d\over dY}\right) O_{R,r,1,0,0,0}&=&
\sqrt{2\over 3}O_{R,r,{1\over 2},-{1\over 2},{1\over 2},{1\over 2}}
+{1\over\sqrt{3}}O_{R,r,{3\over 2},-{1\over 2},{1\over 2},{1\over 2}}\cr
&&+\sqrt{2\over 3}O_{R,r,{1\over 2},-{1\over 2},{1\over 2},{1\over 2}}
+{1\over\sqrt{3}}O_{R,r,{3\over 2},{1\over 2},{1\over 2},-{1\over 2}}
\eea
\bea
{\rm Tr}\left(X{d\over dY}\right) O_{R,r,0,0,0,0}&=&
O_{R,r,{1\over 2},-{1\over 2},{1\over 2},{1\over 2}}
+O_{R,r,{1\over 2},{1\over 2},{1\over 2},-{1\over 2}}
\eea
It is now trivial to verify that
\bea
{\rm Tr}\left(X{d\over dY}\right)
O_{R,r}\left(\begin{gathered}\includegraphics[scale=0.35]{graphss_4}\end{gathered}\right)=0
\eea

The second example we consider illustrates the usual co-product action of the $su(2)_R$ generators.
We will use black edges to denote $Y$ excitations and gray edges to denote $X$ excitations.
Starting from
\bea
O_{R,r}\left(\begin{gathered}\includegraphics[scale=0.45]{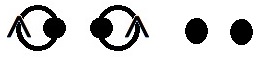}\end{gathered}\right)=
{1\over\sqrt{2}}O_{R,r,1,0,0,0}+{1\over\sqrt{2}}O_{R,r,0,0,0,0}
\eea
and using
\bea
{\rm Tr}\left(X{d\over dY}\right)O_{R,r,1,0,0,0}={1\over\sqrt{2}}
O_{R,r,{1\over 2},-{1\over 2},{1\over 2},{1\over 2}}
+{1\over\sqrt{2}}O_{R,r,{1\over 2},{1\over 2},{1\over 2},-{1\over 2}}
\eea
\bea
{\rm Tr}\left(X{d\over dY}\right)O_{R,r,0,0,0,0}={1\over\sqrt{2}}
O_{R,r,{1\over 2},-{1\over 2},{1\over 2},{1\over 2}}
+{1\over\sqrt{2}}O_{R,r,{1\over 2},{1\over 2},{1\over 2},-{1\over 2}}
\eea
as well as
\bea
O_{R,r}\left(\begin{gathered}\includegraphics[scale=0.45]{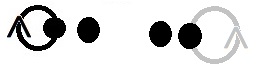}\end{gathered}\right)=
O_{R,r,{1\over 2},-{1\over 2},{1\over 2},{1\over 2}}
\eea
\bea
O_{R,r}\left(\begin{gathered}\includegraphics[scale=0.45]{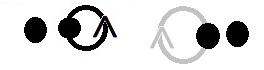}\end{gathered}\right)=
O_{R,r,{1\over 2},{1\over 2},{1\over 2},-{1\over 2}}
\eea
we find
\bea
{\rm Tr}\left(X{d\over dY}\right)
O_{R,r}\left(\begin{gathered}\includegraphics[scale=0.45]{graphss_10}\end{gathered}\right)&=&
O_{R,r}\left(\begin{gathered}\includegraphics[scale=0.45]{graphss_11}\end{gathered}\right)\cr
&&\qquad +
O_{R,r}\left(\begin{gathered}\includegraphics[scale=0.45]{graphss_12}\end{gathered}\right)\cr
&&
\eea
This clearly illustrates that the generator acts on each node individually, turning a black ($Y$) edge into a gray ($X$) edge
when it acts.

In our final example, we would like to test that the coefficient in (\ref{basicRep}) comes out correctly.
Assume that the excitation is built from $j-1$ $Y$ fields and one $X$ field, which all come from the first row of $R$.
In this case we have
\bea
O_{R,r}\left(\begin{gathered}\includegraphics[scale=0.45]{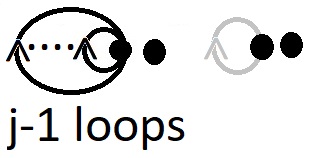}\end{gathered}\right)=
O_{R,r,{j-1\over 2},{j-1\over 2},{1\over 2},{1\over 2}}
\eea
and
\bea
O_{R,r}\left(\begin{gathered}\includegraphics[scale=0.45]{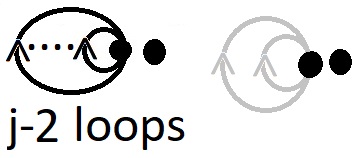}\end{gathered}\right)=
O_{R,r,{j-2\over 2},{j-2\over 2},1,1}
\eea
The equation
\bea
{\rm Tr}\left( X{d\over dY}\right) O_{R,r,{j-1\over 2},{j-1\over 2},{1\over 2},{1\over 2}}
=\sqrt{2(j-1)}O_{R,r,{j-2\over 2},{j-2\over 2},1,1}
\eea
implies
\bea
{\rm Tr}\left( X{d\over dY}\right) O_{R,r}\left(\begin{gathered}\includegraphics[scale=0.45]{graphss_13}\end{gathered}\right)
=\sqrt{2(j-1)}O_{R,r}\left(\begin{gathered}\includegraphics[scale=0.45]{graphss_14}\end{gathered}\right)\cr\cr
\eea
which beautifully matches the expected result of the action of the lowering operator on state $|j,m\rangle$
\bea
R_-|j,j-1\rangle =\sqrt{2(j-1)}|j,j-2\rangle
\eea
A node with $n_Y$ closed $Y$ loops and $n_X$ closed $X$ loops is in the representation $j={1\over 2}(n_Y+n_X)$ and has
$m={1\over 2}(n_Y-n_X)$.

\end{document}